# Thermal conductivity of silicon nanomeshes: Effects of porosity and roughness

Stefanie Wolf[1], Neophytos Neophytou[2*], and Hans Kosina[1]

[1]Institute for Microelectronics, TU Wien, Gußhausstraße 27-29/E360, A-1040 Wien, Austria

[2]School of Engineering, University of Warwick, Coventry, CV4 7AL, UK

*E-mail: N.Neophytou@warwick.ac.uk

## Abstract

We theoretically investigate thermal conductivity in silicon nanomeshes using Monte Carlo simulations of phonon transport. Silicon membranes of 100nm thickness with randomly located pores of 50nm diameter are considered. The effects of material porosity and pore surface roughness are examined. Nanomesh porosity is found to have a strong detrimental effect on thermal conductivity. At room temperature, a porosity of 50% results in ~80% reduction in thermal conductivity. Boundary roughness scattering further degrades thermal conductivity, but its effect is weaker. Thermal transport can additionally be affected by the specific arrangement of the pores along the transport direction.

# I. Introduction

Nanoporous membranes made of single-crystalline Si, also referred to as nanomeshes, are promising candidates for thermoelectric materials due to their extremely low thermal conductivity $\kappa_l$, relatively high thermoelectric power factor, and structure stability that other nanostructured or low-dimensional systems lack [1, 2, 3]. The structure consists of a crystalline Si membrane, usually built on an SOI wafer that includes pores of a certain diameter, pitch, and porosity, either in ordered arrays [1, 2, 3], or disordered [4]. The feature sizes range from 10 to a few 100s of nanometers. Yu *et* al., for example, demonstrated nanomesh structures with large areal footprint of 100μm² with pores of a period of 34nm and diameters of 11nm and 16nm in a 22nm thin crystalline Si film on an SOI wafer [1]. Tang *et* al. [2] have fabricated a highly dense hexagonal pore pattern of pitch down to 55nm with 35% porosity. Both works reported thermal conductivities of 1-2W/mK, which is two orders of magnitude smaller than the bulk Si value, close to the amorphous silica limit. Similar numbers have also been reported for sintered porous films [4]. Hopkins *et* al. [3] demonstrated even larger footprints, of the order of 10,000 μm², but with holes of diameters 300nm-400nm and spacing of 500nm-800nm, and also reported very low thermal conductivites. With respect to their thermoelectric performance, a *ZT* value of 0.4 was reported at 300K, a large increase compared to the Silicon bulk value $ZT_{bulk}$ ~ 0.01, and one of the best values reported for Si [2]. Such *ZT* values are very comparable to those reported for Si nanowires in Refs [5, 6]. Nanomeshes, however, are advantageous in terms of mechanical stability and the feasibility of large scale production from a manufacturing point of view [7].

In such geometries phonons encounter enhanced boundary scattering on the nanopore sidewalls, which results in the extremely low thermal conductivities observed. This was also shown theoretically for Si as well as SiGe periodic nanomeshes with small pore diameters up to 3nm using molecular dynamics simulations [8, 9, 10]. For larger pores with diameters of tens to hundreds of nanometers several works use continuum analytical models, or the Boltzmann Transport Equation (BTE) for phonon transport assuming a periodic structure, which simplify the simulation complexity significantly [11, 12, 13, 14, 15].

Another approach to study thermal transport in such materials is to numerically solve the BTE for phonons using the Monte Carlo (MC) method. This approach has been successfully applied to bulk materials [16], thin films [17, 18], nanowires [19, 20, 21, 22], nano-composites [23], and polycrystalline materials [24], and allows for the treatment of arbitrary channel geometries. It is most suitable for describing phonon transport in large scale nanomeshes and nanoporous materials in which the pores are not necessarily periodically arranged, or when the external boundaries of the nanomesh contribute significantly to phonon scattering. At present such studies are not available, although these geometries are attractive for various applications. In this work, therefore, we calculate the thermal conductivity of nanoporous single crystalline Si membranes with randomly placed pores by numerically solving the BTE for phonons using a *single-phonon* MC method. This method involves features which are slightly different compared to the common multi-phonon methods, and we explain it in detail in the 'Approach' section. Our goal is to elucidate the influence of three parameters on the thermal conductivity: i) the porosity, ii) the roughness amplitude of the pore and the membrane surfaces, and iii) the influence of spatial variations in the pore placement on the thermal conductivity. The findings we present below deliver insight into the transport of phonons in materials that include randomized scatterers along the phonon paths such as nanomeshes and materials with embedded nano-inclusions.

## II. Approach

The simulated devices consist of the channel domain containing the randomly placed pores, and two thermal contacts acting as a heat source and a heat sink with a black body spectrum. The device we consider is 100nm thick, 500nm wide, and 1μm long. To simulate the device geometry we use a tetrahedral simulation grid generated by the *Global TCAD Solutions* simulation framework [25]. The MC algorithm accounts for longitudinal and transversal acoustic phonons and nonlinear dispersion relations. Following the work of Ref. [26], we use the expression $\omega(k) = v_s k + ck^2$ to fit the bulk dispersion relation of phonons under the isotropic Brillouin zone approximation, where $k$ is the norm of the wave vector. The use of bulk phonon dispersion was shown to be a valid approximation for describing phonon transport in nanostructures of feature sizes down to several nanometers at room temperature [27]. The phonon bands could change significantly at the nanoscale, but this is the case for channels with ultra-thin feature sizes [28], which in this work we do not consider. We include the longitudinal acoustic (LA) and transverse acoustic (TA) phonon branches. The dispersion coefficients are $v_s$ = 9.01 x $10^3$ m/s and c=-2 x $10^{-7}$ m$^2$/s for the LA branch, and $v_s$ = 5.23 x $10^3$ m/s and c=-2.26 x $10^{-7}$ m$^2$/s for the TA branch [26].

The simulation steps are as follows:

The temperature of the thermal contacts is set to the constant values $T_H$ (hot) and $T_C$ (cold). Initially, the temperature of the cells in the actual simulation domain is also set to $T_C$. Phonons are then initialized. *A single phonon at a time* is generated at a random position in the thermal contacts according to their temperature (most at the hot contact) and injected into the simulation domain. The following quantities are initialized by following the phonon Monte Carlo schemes described in Refs. [16, 17, 22]:

*Phonon position:* The position vector $\vec{r}$ of any point within a cell is calculated using $\vec{r} = \vec{r}_A + \vec{b} R_1 + \vec{c} R_2 + \vec{d} R_3$ where $\vec{r}_A$ is the position vector of one of the grid cell tetrahedron's vertex $A$, and $\vec{b}$, $\vec{c}$, and $\vec{d}$ are the edge-vectors from $A$ to all other vertices. $R_1$, $R_2$ and $R_3$ are random numbers such that $\sum_i R_i < 1$.

*Phonon frequency:* We separate the frequency spectrum in $N_\omega$=1000 equidistant spectral intervals of width $\Delta_\omega$. The number of phonons in the $i$-th spectral interval is given by $N_i = \left[ g_{LA} D_{LA}(\omega_i) + g_{TA} D_{TA}(\omega_i) \right] \langle n(\omega_i) \rangle \Delta \omega_i$, where $\langle n(\omega_i) \rangle$ represents the equilibrium Boltzmann distribution, and $D$ and $g$ are the density of states and degeneracy of the LA and TA phonon branches, respectively. The normalized cumulative number density function is given by:

$$F_i = \frac{\sum_{k=1}^{i} N_k}{\sum_{k=1}^{N_\omega} N_k}. \tag{1}$$

Then a random number $R$ between zero and unity is drawn. If $F_{i-1} < R \leq F_i$ the phonon belongs to the $i$-th spectral interval and the phonon frequency is given by $\omega = \omega_i + (R-1)\Delta\omega$.

*Phonon Polarization:* The probability of a phonon being in the $i$-th spectral interval and having LA polarization is expressed as:

$$P_i^{LA} = \frac{N_i^{LA}}{N_i^{LA} + N_i^{TA}}. \tag{2}$$

If a drawn random number $R$ is less than $P_i^{LA}$ the phonon belongs to the LA branch. Otherwise it belongs to a TA branch.

*Phonon Wave Vector:* The norm of the wave vector $k$ can be determined directly by inversion of the dispersion relationship. The direction of the wave vector is given by $\hat{k} = \hat{e}_x \sin\vartheta \cos\psi + \hat{e}_y \sin\vartheta \sin\psi + \hat{e}_z \cos\vartheta$, where $\vartheta$ and $\psi$ are random angles.

*Group Velocity:* The group velocity is defined as the slope of the dispersion being $v_g = v_s + 2ck$ and is parallel to the wave vector.

The initialized phonon then performs a free flight of duration $\Delta t$. At the end of the free flight, the phonon has a new position, and could scatter (see Fig. 1a). If a phonon reaches a device boundary at the top or bottom of the channel or a pore, we then have a boundary scattering event. Boundary scattering changes the phonon direction, but not its frequency. For boundary scattering we use a constant specularity parameter p, ranging from 0 to 1, to describe diffusive (p=0), or specular (p=1) boundary scattering. For each boundary scattering event a random number $0<R<1$ is drawn. If $R>p$, then the phonon direction is randomly reset. Otherwise the phonon is mirror-like reflected. Phonon-phonon scattering is treated using frequency- and temperature-dependent phonon lifetimes for both polarization branches [29]. In the phonon Monte Carlo method for Si, phonon-phonon scattering is commonly described as [16, 17]:

$$\tau_{LA}^{-1} = B_{NU}^{LA}\omega^2 T^3 \qquad (3a)$$

$$\tau_{TA,N}^{-1} = B_{N}^{TA}\omega T^4 \qquad (3b)$$

$$\tau_{TA,U}^{-1} = \begin{cases} 0 & \text{for } \omega < \omega_{1/2} \\ B_{U}^{TA}\omega^2 / \sinh\left(\frac{\hbar\omega}{k_B T}\right) & \text{for } \omega \geq \omega_{1/2} \end{cases} \qquad (3c)$$

Here N represents the normal and U the Umklapp (three-phonon) processes, $\omega_{1/2}$ is the frequency corresponding to $k = k_{max}/2$, $B_{NU}^{LA} = 2\times10^{-24}\,s/K^3$, $B_{N}^{TA} = 9.3\times10^{-13}K^{-4}$ and $B_{U}^{TA} = 5.5\times10^{-18}s$. After a three-phonon scattering event the phonon frequency and direction are reset, whereas after a normal process the frequency and the magnitudes of the wavevector and velocity are reset, following the scheme described in Ref. [16].

After every free flight time step the energy and temperature of the cells is calculated. To determine the cell temperature, we compute the total phonon energy in each cell, given by:

$$E = \frac{V}{W}\sum_{p}\sum_{i}\frac{\hbar\omega_i}{\exp\left(\frac{\hbar\omega_i}{k_B T}\right)-1}g_p\frac{k_i^2}{2\pi v_i^2}\Delta\omega_i, \qquad (4)$$

where $V$ is the cell volume, $i$ is the spectral interval index, $p$ is the polarization index, $k$ is the wave number and $g_p$ the degeneracy of the dispersion branches. Once the energy of each cell is known, the temperature is calculated by numerical inversion of Eq. 4 using the Newton-Raphson method. In Eq. 4, a scaling factor $W$ is used as commonly employed in phonon MC methods [16, 17, 22]. $W$ is the number of phonons corresponding to one simulated phonon. Scaling is needed because the mean cell energy is many orders of magnitude larger than a single phonon's energy. The number of phonons per unit volume is determined by the Bose-Einstein distribution and the phonon density of states $D$ as:

$$N = \sum_{p}\sum_{i}\frac{1}{\exp\left(\frac{\hbar\omega_i}{k_B T}\right)-1}D(\omega_i,p)\Delta\omega_i \qquad (5)$$

This can be a very large number, in particular in Si at room temperature there are about ~$10^5$ phonons even in a small volume as (10nm)$^3$ [22]. Simulation of such large number of phonons takes too much computation time. To reduce the number of simulated phonons to *N\**, a commonly employed method is to scale the number of phonons by a scaling factor *W* as *N\*=N/W*, which means that each stochastic sample used during the simulation actually represents an ensemble (or bundle) of *W* phonons [30]. In our case, for room temperature, we use $W=10^5$, a similar number used in previous studies as well [16].

The simulation begins by initializing one phonon at a time in the contacts and injecting it into the channel. The injected phonon drifts in the channel. When the phonon enters a new cell: i) the phonon energy is added to the cell's energy, ii) the new cell temperature is calculated by inversion of Eq. 4, and iii) a scattering event could happen according to a certain probability, and energy could be dissipated or absorbed. In the next free flight phase the phonon leaves the cell and its energy is subtracted from that cell. Therefore, during consequent time steps *Δt*, the energy and temperature of the channel cells change when a phonon drifts into a particular cell, leaves the cell, or undergoes phonon-phonon scattering in that cell. Since phonon-phonon scattering changes the phonon frequency, when updating the phonon energy, we add the energy difference between the old and new phonon to the energy of the simulation domain cell in which the scattering event has taken place as $E_{cell} \rightarrow E_{cell} + \hbar\Delta\omega$. This ensures energy conservation. After the scattering event, the simulation is continued by tracking the drift of the new phonon (with its new properties) from the position where the scattering event occurred. The drift steps are repeated until the phonon finally reaches a contact where it gets absorbed. Then the simulation continues with a newly created phonon.

During the drift, the phonon energy as well as the energy and temperature of the cells in which it scatters, can change as follows: If a low energy phonon undergoes a three-phonon scattering event in a hot cell, the phonon will most probably gain energy, and the temperature of the cell will decrease. In the opposite situation, if a high energy phonon scatters in a cold cell, the phonon will most probably be annihilated and the temperature will rise. As a result, since the cells away from the heat source are at lower temperature than the heat source temperature $T_H$, the temperature gradually increases along the phonon transport direction. This sequence is repeated until a steady state is reached, in which case a linear temperature drop from the hot to the cold contacts is established, as shown in Fig. 1b. Steady state is reached when the temperature of the cells in the channel does not change (according to an error criterion) within 10,000 single phonon initializations/injections. We typically need $8 \times 10^7$ phonons to reach at this state.

After steady state is reached, we calculate the heat flux by sequentially injecting a prescribed number of phonons ($N$) first from the source, ("H" in Fig. 1a), into the device. The typical number of phonons used is $N \sim 4 \times 10^6$. Again, this is done a single phonon at a time. We calculate the sum $E_{in}^{H}$ of the phonons incident energies. The energy of all back-scattered phonons (leaving the device back from the source contact) is summed up to $E_{out}^{H}$. Additionally, we calculate the average time it takes for a phonon to travel the distance through the device (average time-of-flight, $<TOF>$). Then we repeat the procedure by initializing the same number of phonons $N$ from the heat sink (cold side, "C" in Fig. 1a) and calculate $E_{in}^{C}$ and $E_{out}^{C}$. The total phonon flux is then given by:

$$\Phi = N_{tot} \frac{\left(E_{in}^{H} - E_{out}^{H}\right) - (E_{in}^{C} - E_{out}^{C})}{N <TOF>}, \qquad (6)$$

where $N_{tot}$ is the total number of phonons inside the device, as dictated by the Bose-Einstein distribution and the local temperature. The idea here is that each phonon in the domain of the channel contributes to a certain net energy flow per unit time (a net energy rate per phonon) through the volume of the channel. This is computed by the MC injection scheme. The total flux is found by multiplying this average rate with the actual total number of phonons in the volume. Finally, the thermal conductivity is calculated using the heat flux through the medium for a given temperature difference $\Delta T$ by applying Fourier's law as $\kappa = -\Phi/\Delta T$.

Note that the single-phonon approach we follow differs from the multi-phonon approaches described in the literature. In the multi-phonon approaches, a certain number of phonons are initialized at once according to the Bose-Einstein distribution and injected in the channel. The trajectories of all phonons are traced simultaneously at every time step $\Delta t$ and the energy and temperature of all cells are also updated every $\Delta t$. The net flux upon convergence is computed at any interface perpendicular to the transport direction simply by calculating the net rate at which phonons cross that interface. In our single phonon approach, however, since we do not have access to all phonons travelling in the channel at all times, the flux is calculated by determining the average flux due to one phonon crossing the entire domain, and then multiplying by the total number of phonons, as described above. The main difference, however, comes from the fact that we keep track of the trajectory of one phonon at a time, from initialization to thermalization. The path of each phonon, therefore, is independent of the other phonons, but every subsequent phonon feels the full effect of the previous phonons on the cells of the simulated channel. While initially (before steady state) this approach does not correspond to the real physical situation in which multiple phonons travel simultaneously across the channel, after a large number of single phonon trajectories are calculated the simulation indeed reaches a steady state and the a linear temperature gradient is finally build across the channel (see Fig. 1b). In our single-particle approach, thus, the statistics are accumulated by repetition of the simulation sequence with a single particle. The simulator was also validated by providing very good agreement with the thermal conductivity of bulk Si over a variety of temperatures. The main advantage of the single-particle scheme is that it requires much less memory as it keeps track of only one phonon at a time, which is beneficial in complex structures as the ones we consider, where the number of cells $N_{cells}$ can reach very large numbers.

Figures 1b and 1c show an example of a simulated device geometry. Figure 1b shows the temperature gradient, which is almost linear from the hot to the cold end of the channel (in this particular case $T_H=310K$ and $T_C=290K$), which indicates that the single phonon injection scheme works properly. Figure 1c shows a few of the phonon trajectories propagating through the channel.

## III. Results and Discussion

The data in Fig. 2 summarizes the simulation results for the thermal conductivity in the nanomeshes examined. We examine the influence of the porosity $\Phi$, and of the boundary roughness amplitude characterized by the specularity parameter p. The thermal conductivity is plotted as a function of porosity $\Phi$ (x-axis), and results for structures with different boundary specularity parameters are shown (from p=1 for fully specular boundaries to p=0.1 for almost fully diffusive boundaries). Each data point is computed by taking the average of 100 channels with different randomized pore arrangements. The standard deviation of their distribution is denoted by the error bars. The pore diameter is fixed at 50 nm.

Three main conclusions can be deduced from this figure: i) As expected, the thermal conductivity decreases with increasing porosity. Increasing the porosity from 0% to 35%, results in a three-fold reduction in thermal conductivity. ii) Under the same porosity conditions, increasing the roughness strength by an order of magnitude (from p=1 down to p=0.1), reduces the thermal conductivity by only ~40% (see the difference between the upper and lower curves in Fig. 2). Under the same roughness conditions, however, it takes only 10-20% increase in the porosity to reduce the thermal conductivity by the same amount (~40%). iii) Roughness with p=0.5 results in a significant reduction in thermal conductivity. The influence on the thermal conductivity of additional increases in the roughness strength is, however, weak.

Experimental measurements for the thermal conductivity of nanomeshes are also indicated in Fig. 2. The data are taken from Refs. [2, 3, 4]. Direct comparison between measurements and our simulations is not possible because although we compare at similar porosities, the pore diameters and pitch are very different. A good agreement is observed between our calculations and measurements for sintered porous Si from Ref. [4], and some of the results of Ref. [2]. Several measured points, however, are lower than our calculations. This is an indication that other thermal conductivity degrading mechanisms can exist in the fabricated nanomeshes, i.e. phonon scattering through defects, dopants, or impurities. Another important mechanism is the possible existence of phonon coherent effects, in which case the phonon spectrum is altered. The materials of Refs [2] and [3] are ordered, periodic nanomeshes that form phononic lattices, in which case coherent events could even produce bandgaps in the phonon spectrum that significantly reduce the thermal conductivity. Our semiclassical results for randomized pore meshes, therefore, indirectly indicate the existence of coherent phonon effects in the ordered nanomeshes and quantify the amount of additional thermal reduction that can be achieved.

Figure 3 highlights the influence of *boundary roughness scattering* of phonons on the pore surfaces. Figure 3a shows the distributions of the thermal conductivities of simulated devices of porosity $\Phi=15\%$, while the roughness amplitude of the pore surfaces changes from completely specular (blue lines, p=1) to 90% diffusive (orange lines, p=0.1). As the roughness strength increases, the thermal conductivity decreases, (i.e. it shifts from $\kappa_l$~50W/mK down to $\kappa_l$~32W/mK). It takes, however, only moderate roughness amplitude to achieve most of the reduction of the thermal conductivity. As p decreases from p=1 down to p=0.5 (green line), a stronger reduction is observed. When further doubling the roughness strength by changing p from p=0.5 to p=0.25 (red line) and even down to p=0.1 (orange line), the thermal conductivity drops only marginally. Figure 3b shows the same results as Fig. 3a, but this time the devices considered are of larger porosity $\Phi=50\%$. In this case, the larger porosity further enhances the effects we described above for Fig. 3a. An initial drop in the thermal conductivity when p is reduced from p=1 (blue line) to p=0.5 (green line) is observed. A very small change is observed, however, as p is reduced even further down to p=0.25 (red line). A reduction from p=0.25 to p=0.1 (orange line) leaves the thermal conductivity almost unaffected. It is also evident that the distributions become narrower as the roughness increases, although the normalized spread in the distribution to the actual thermal conductivity is constant (see inset of Fig. 3b). The normalized standard deviations of the distributions are larger for larger porosity since the existence of more pores creates more randomness. An important message conveyed here is that very strong roughness in nanoporous materials does not necessarily have a much larger influence on the thermal conductivity compared to moderate roughness. At large porosities, especially, due to the larger relative spread in the thermal conductivity distributions, it seems that the strength of boundary scattering is irrelevant. At large porosities the phonon flow is almost completely randomized by the large number of pores and boundaries, and any further randomization by the morphology of the boundary roughness (for very low p) is redundant. This indicates again the relative importance of the porosity over roughness. For thermoelectrics this can be a positive effect, since strong roughness reduces the electronic conductivity and should be avoided. The nature of the boundary roughness and its influence in randomizing phonon trajectories seems to be somewhat more influential at smaller porosities (<30%).

In Fig. 4 we now illustrate the *influence of porosity Φ* in more detail. Figure 4a shows the distributions of the thermal conductivities of the simulated devices with perfect boundaries (p=1), while the porosity changes from Φ=15% (blue line) to 50% (orange line). As the porosity increases, the thermal conductivity decreases from $\kappa_l$~50W/mK down to $\kappa_l$ ~15W/mK. The observation is the same in the case of channels with rough pores (p=0.1), as shown in Fig. 4b. Of course the thermal conductivity in this diffusive boundary case is lower compared to the specular boundary devices of Fig. 4a. Interestingly, the increase in porosity degrades the thermal conductivity almost linearly. As the porosity increases, a noticeable thermal conductivity reduction is also observed at every step independently of the pore roughness. The consistent and linear reduction in the thermal conductivity is in contrast to the influence of boundary roughness described in Fig. 3, in which case the thermal conductivity depends on the roughness magnitude very weakly for p<0.5. The reason that the thermal conductivity decreases noticeably when the porosity increases, is that the porosity decreases the pathways for phonon transport, which increases the thermal resistance of the material.

We note that the pore diameters we consider in this work are in all cases 50nm. The average distance between the pores (from the circumference of one pore to the circumference of the nearest pore) is determined by the porosity of the material, and extends roughly from ~60nm for geometries of large porosities (Φ=50%) to ~115nm for geometries of small porosities (Φ=15%). If the separation of the pores is smaller than the phonon-phonon scattering mean-free-path (MFP), boundary scattering dominates the thermal conductivity. In the opposite limit, if the separation is much larger than the phonon-phonon scattering MFP, the internal three-phonon scattering dominates and the classical limit is recovered [11]. This interplay between internal and interface scattering is also observed and well described by Melis *et* al. for nanocomposites [31] and Aksamija *et* al. for superlattice structures [32]. It is, therefore, useful to examine at which regime the nanomesh geometries we consider operate. As shown by Jeong *et* al., in bulk Si more than 50% of the heat is carried by phonons with MFPs longer than 1μm, with the average phonon MFP at room temperature being ~132nm [33]. In thin films 50% of the heat is also carried by phonons with MFPs larger than the film thickness [33], whereas long MFP phonons dominate transport in Si nanowires as well [34]. For the nanomesh geometries we examine, for large porosities the distances between the pores are smaller than the dominant phonon MFPs, and our simulations are actually at the limit where three-phonon internal scattering is weak compared to boundary scattering. In this case, by reducing the porosity the thermal conductivity increases almost linearly, as observed in Fig. 2 and Fig. 4. Once the porosity reaches very small values, which means that the distances between the pores are large, even larger than the phonon-phonon scattering MFPs, the thermal conductivity should saturate to the classical limit. From Fig. 2, however, the thermal conductivity does not saturate even at 15% porosities, the smallest ones we consider. In the case of specular boundaries ($p=1$), the thermal conductivity even increases more than linearly with porosity decrease. It is, therefore, evident that in all of our simulations the structures are dominated by boundary scattering rather than internal phonon-phonon scattering. Internal scattering could start to influence the thermal conductivity once the porosity of the material is reduced well below 10%, in which case the pore distance will extend beyond 100-200nm, and will be larger than the average phonon MFP. Then, the influence of roughness and even the presence of pores will be less significant. This work, therefore, specifically describes how the presence and density of pore boundaries affects heat flow, and how the roughness of those boundaries adds to that. Although three-phonon scattering is included in the calculation, it is a weaker mechanism than boundary scattering and does not influence our conclusions.

An additional factor that affects the sensitivity of the thermal conductivity is the *placement of the pores* in the structures from segment to segment along the heat flow direction, and the number of pores in each segment. Figure 5a and 5b illustrate two example structures with two different situations of pore arrangement within segments along the transport direction. In either case the porosity is the same (25 pores per channel). In Fig. 5a, the pores are uniformly distributed along the segments of the material. In Fig. 5b, however, the pore density significantly varies from segment to segment. In the second and fourth segments of the material, the pore density is high, whereas in the rest of the segments the density is low. Although this illustration depicts two extreme cases, such variations in pore arrangement (of course smaller) exist, and will affect the thermal conductivity. To quantify this effect in our structures, we separate the channel of the simulated materials in segments of length 100nm along the transport direction. We then measure the number of pores within each segment, and compute the standard deviation of this distribution $\sigma_h$. For example, the geometry of Fig. 5b will have larger $\sigma_h$ than the geometry of Fig. 5a. Figure 5c shows the thermal conductivity of 100 simulated channels of 50% porosity versus $\sigma_h$. For each of the 100 devices with different pore arrangements we compute the thermal conductivity assuming boundary specularity parameters, p=1 (blue triangles), p=0.5 (green squares), p=0.25 (red triangles), p=0.1 (orange circles). Figure 5c, thus, depicts results for 400 simulated devices. The lines indicate a linear fit for the thermal conductivities. Two main observations can be extracted: i) The thermal conductivity is a sensitive function of the exact pore arrangement, which is indicated by the spread in the data points. Indeed, the correlation coefficient of the linear least squares fit (lines in Fig. 5 for each p) is rather small, ranging from $r$=0.224 for $p$=0.1, to $r$=0.269 for p=1. ii) The thermal conductivity on average decreases as the standard deviation of the pores per slice increases. This indicates an increase in the local thermal resistance along the transport direction, which originates from the segments which contain more pores. And as the pore deviation per segment increases, it means that there exist more segments with higher concentration of pores compared to the average pore concentration per segment. Of course segments with low pore concentration also exist in the same channel (since we compare at the same porosity). The increase in thermal resistance due to the pore dense segments seems to influence phonon transport more, and the thermal conductivity is overall reduced. The larger relative thermal conductivity reduction is observed for specular boundaries where p=1, and it is of the order of ~40%.

## IV. Conclusions

In summary, in this work we have investigated the thermal conductivity in silicon nanomeshes using Monte Carlo simulations for phonons. A single-phonon approach is used and presented in detail, in which phonons are simulated one at a time. We show that the nanomesh porosity has a strong detrimental influence on the thermal conductivity. Boundary roughness still degrades the thermal conductivity, but its influence is smaller, especially in materials with porosity Φ>35%. Interestingly, very strong pore roughness does not necessarily have a stronger influence on the thermal conductivity compared to moderate pore roughness. We show that the relative spread in the distribution of the thermal conductivity of the 400 structures which were simulated for each porosity increases with increasing porosity, but it is rather unaffected by the roughness. Finally, we demonstrate that the specific arrangement of pores from segment to segment along the transport direction also affects thermal transport. As the variation in the number of pores from segment to segment increases, the thermal conductivity degrades noticeably.

## Acknowledgements

This work is supported by the Austrian Science Fund (FWF) contract P25368-N30. The authors acknowledge helpful discussions with Dr. Hossein Karamitaheri and Zlatan Stanojevic.

# References

[1]. J.-K. Yu, S. Mitrovic, D. Tham, J. Varghese and J. R. Heath, *Reduction of Thermal Conductivity in Phononic Nanomesh Structures,* Nature Nanotechnology, 5 (10), 718-721, 2010

[2] J. Tang, H.-T. Wang, D. H. Lee, M. Fardy, Z. Huo, T. P. Russell and P. Yang, *Holey Silicon as an Efficient Thermoelectric Material*, Nano Letters, Vol. 10(10), 4279-4283, 2010

[3] P. E. Hopkins, C. M. Reinke, M. F. Su, R. H. Olsson III, E. A. Shaner, Z. C. Leseman, J. R. Serrano, L. M. Phinney, and I. E. Kady, *"Reduction in the Thermal Conductivity of Single Crystalline Silicon by Phononic Crystal Patterning,"* Nano Lett., 11, 1, 107-112, 2011

[4] A. Wolf and R. Brendel, *Thermal conductivity of sintered porous silicon films,* This Solid Films, 513, 385, 2006

[5] A. I. Boukai, Y. Bunimovich, J. Tahir-Kheli, J. K. Yu, W. A. Goddard, J. R. Heath, *Silicon nanowires as efficient thermoelectric materials*, Nature 451(7175), 168-171, 2008

[6] A. I. Hochbaum, R. Chen, R. D. Delgado, W. Liang, E. C. Garnett, M. Najarian, A. Majumdar and P. Yang, *Enhanced thermoelectric performance of rough silicon nanowires*, Nature 451, 163-167, 2008

[7] K. Nielsch, J. Bachmann, J. Kimling, and H. Bottner, *Thermoelectric Nanostructures: From Physical Model Systems towards Nanograined Composites,* Adv. Energy Mater., 1, 713-731, 2011

[8] J.-H. Lee, G. Galli, and J. C. Grossman, *Nanoporous Si as an Efficient Thermoelectric Material,* Nano Lett., 8, 11, 3750-3754, 2008

[9] S. P. Hepplestone and G. P. Srivastava, *Lattice dynamics and thermal properties of phononic semiconductors,* Phys. Rev. B, 84, 115326, 2011

[10] C. Bera, N. Mingo and S. Volz, *Marked Effects of Alloying on the Thermal Conductivity of Nanoporous Materials*, Phys. Rev. Lett., 104, 115502, 2010

[11] R. Prasher, *Transverse thermal conductivity of porous materials made from aligned nano- and microcylindrical pores*, J. Appl. Physics, 100, 064302, 2006.

[12] Q. Hao, G. Chen and M. S. Jeng, *Frequency-dependent Monte Carlo simulations of phonon transport in two-dimensional porous silicon with aligned pores*, Journal of Applied Physics, 106(11), 114321-114321, 2009

[13] J. D. Chung and M. Kaviany, *Effects of Phonon Pore Scattering and Pore Randomness on Effective Conductivity of Porous Silicon*, International Journal of Heat and Mass Transfer, Vol. 43(4), 521-538, 2000

[14] K. Miyazaki, T. Arashi, D. Makino and H. Tsukamoto, *Heat Conduction in Microstructured Materials*, IEEE Transactions on Components and Packaging Technologies, Vol. 29(2), 247-253, 2006

[15] R. H. Tarkhanyana and D. G. Niarchos, *Reduction in lattice thermal conductivity of porous materials due to inhomogeneous porosity*, International Journal of Thermal Sciences, 67, 107–112, 2013

[16] D. Lacroix, K. Joulain and D. Lemonnier, *Monte Carlo transient phonon transport in silicon and germanium at nanoscales*, Phys. Rev. B 72, 064305, 2005

[17] S. Mazumder and A. Majumdar, *Monte Carlo Study of Phonon Transport in Solid Thin Films Including Dispersion and Polarization*, J. Heat Transfer 123, 749, 2001

[18] Z. Aksamija and I. Knezevic, *Anisotropy and boundary scattering in the lattice thermal conductivity of silicon nanomembranes*, Phys. Rev. B, 82, 045319, 2010

[19] Y. Chen, D. Li, J. R. Lukes, and A. Majumdar, *Monte Carlo simulation of silicon nanowire thermal conductivity*, J. Heat Transfer, 127, 1129, 2005

[20] D. Lacroix, K. Joulain, D. Terris, and D. Lemonnier, *Monte Carlo Simulation of Phonon Confinement in Silicon Nanostructures: Application to the Determination of the Thermal Conductivity of Silicon Nanowires*, Applied Physics Letters, 89(10), 103104, 2006

[21] J. Randrianalisoa and D. Baillis, *Monte Carlo Simulation of Steady-State Microscale Phonon Heat Transport*, Journal of Heat Transfer, 130(7), 072404, 2008

[22] E. B. Ramayya, L. N. Maurer, A. H. Davoody, and I. Knezevic, *Thermoelectric properties of ultrathin silicon nanowires*, Phys. Rev. B, 86, 115328, 2012.

[23] M.-S, Jeng, D. Song, G. Chen and R. Yang, *Modeling the Thermal Conductivity and Phonon Transport in Nanoparticle Composites Using Monte Carlo Simulation,* J. Heat Transfer 130(4), 042410, 2008

[24] A. McGaughey and A. Jain, *Nanostructure thermal conductivity prediction by Monte Carlo sampling of phonon free paths*, Appl. Phys. Lett., 100, 061911, 2012.

[25] http://www.globaltcad.com/en/home.html

[26] E. Pop, R. W. Dutton and K. E. Goodson, *Analytic band Monte Carlo model for electron transport in Si including acoustic and optical phonon dispersion*, J. Appl. Phys 96, 4998, 2004

[27] R. Prasher, T. Tong, and A. Majumdar, *Approximate Analytical Models for Phonon Specific Heat and Ballistic Thermal Conductance of Nanowires*, Nano Lett., 8, 99, 2008.

[28] H. Karamitaheri, N. Neophytou, M. K. Taheri, R. Faez, and H. Kosina, *Calculation of Confined Phonon Spectrum in Narrow Silicon Nanowires using the Valence Force Field Method,* J. Electronic Materials, 42, 7, 2091-2097, 2013.

[29] M. G. Holland, *Analysis of Lattice Thermal Conductivity*, Phys. Rev., 132, 6, 2461-2471, 1963

[30] R. Peterson, J. Heat Transfer, *Direct Simulation of Phonon-Mediated Heat Transfer in a Debye Crystal,* 116, 815, 1994.

[31] C. Melis and L. Colombo, *Lattice thermal conductivity of $Si_{1-x}Ge_x$ nanocomposites*, Phys. Rev. Lett., 112, 065901, 2014.

[32] Z. Aksamija and I. Knezevic, *Thermal conductivity of $Si_{1-x}Ge_x/Si_{1-y}Ge_y$ superlattices: Competition between interfacial and internal scattering.* Phys. Rev. B, 88, 155318, 2013.

[33] C. Jeong, S. Datta, and M. Lundstrom, *Thermal conductivity of bulk and thin-film silicon: A Landauer approach*, J. Appl. Physics, 111, 093708, 2012.

[34] H. Karamitaheri, N. Neophytou, and H. Kosina, *Anomalous diameter dependence of thermal transport in ultra-narrow Si nanowires*, J. Appl. Phys. 115, 024302, 2014.

Figure 1:

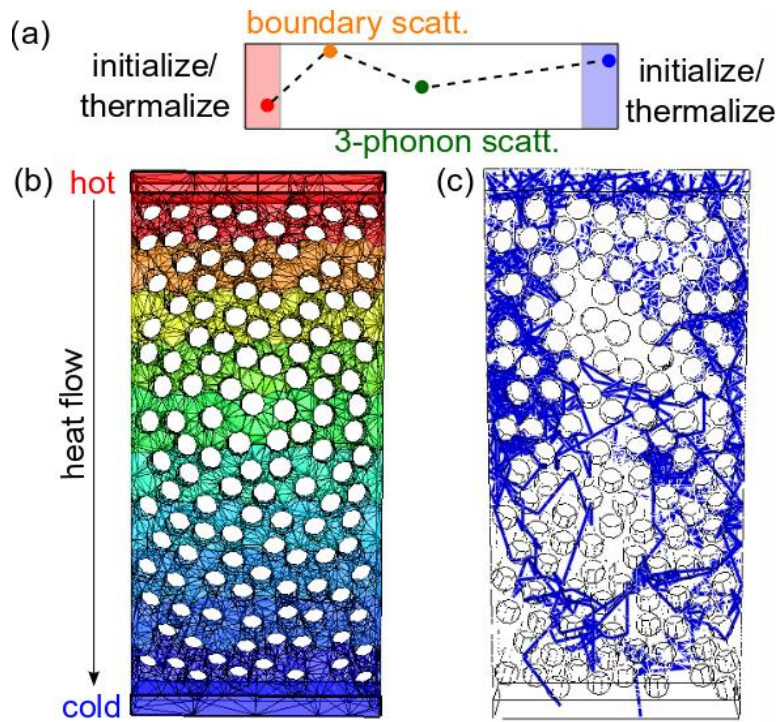

Figure 1 caption:

(a) Schematic of the phonon Monte Carlo simulation scheme. Phonons are initialized at both (ideal) contacts, and travel through the channel, undergoing Umklapp and boundary scattering. When they reach the contacts they are thermalized. (b) Nanoporous device geometry (with random pore arrangement of 50m in diameter). The colormap indicates the calculated temperature gradient (linear drop). Hot contact temperature $T_H$=310K and cold contact temperature $T_C$=290K are assumed. (c) Phonon trajectories in the nanoporous geometry. Here the phonons are initialized at the top contact.

Figure 2:

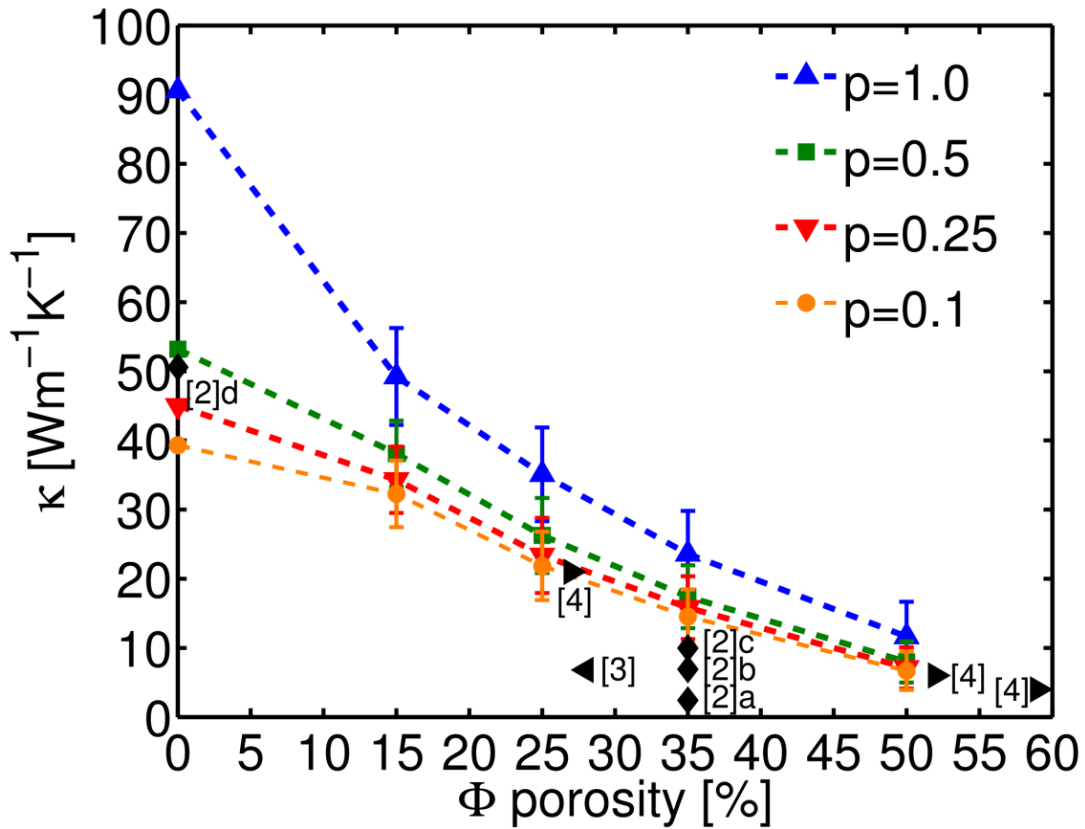

Figure 2 caption:

The thermal conductivity versus material porosity. Calculations: Cases for pore surface roughness specularity parameter p=1 (specular), p=0.5, p=0.25, and p=0.1 (90% diffusive) are shown. Each data point represents the average of 100 structure samples, whereas the error bar denotes the standard deviation of the thermal conductivity of the samples. Experimental data for similar porosities from references [2], [3], and [4] are indicated (pitch and pore diameters can differ—i.e., the structures [2]a–c, [3]).

Figure 3:

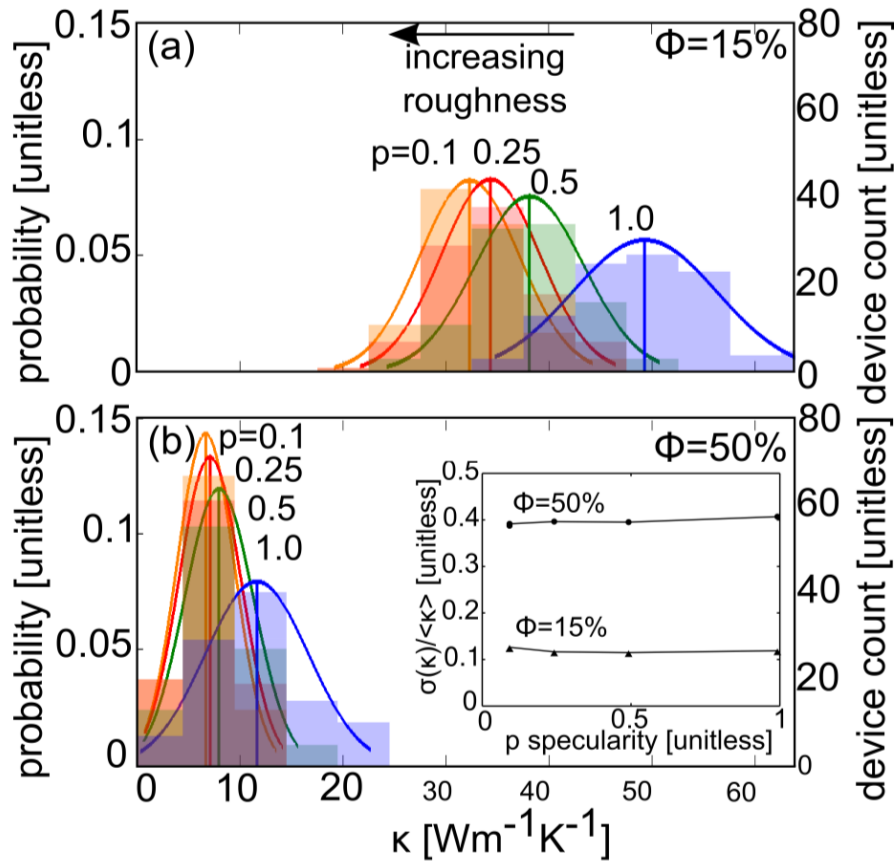

Figure 3 caption:

The thermal conductivity distributions in the nanomeshes for different boundary roughness conditions. (a) Results for porosity Φ=15%. (b) Results for porosity Φ=50%. Results for phonon-boundary scattering specularity parameter p=1 (blue), p=0.5 (green), p=0.25 (red), and p=0.1 (orange) are shown. 100 devices are simulated for each curve. The device count is indicated in the right y-axis. The arrow indicates the direction of increasing roughness. Inset: The relative standard deviation of the distributions versus specularity p for porosities Φ=50% and Φ=15%.

Figure 4:

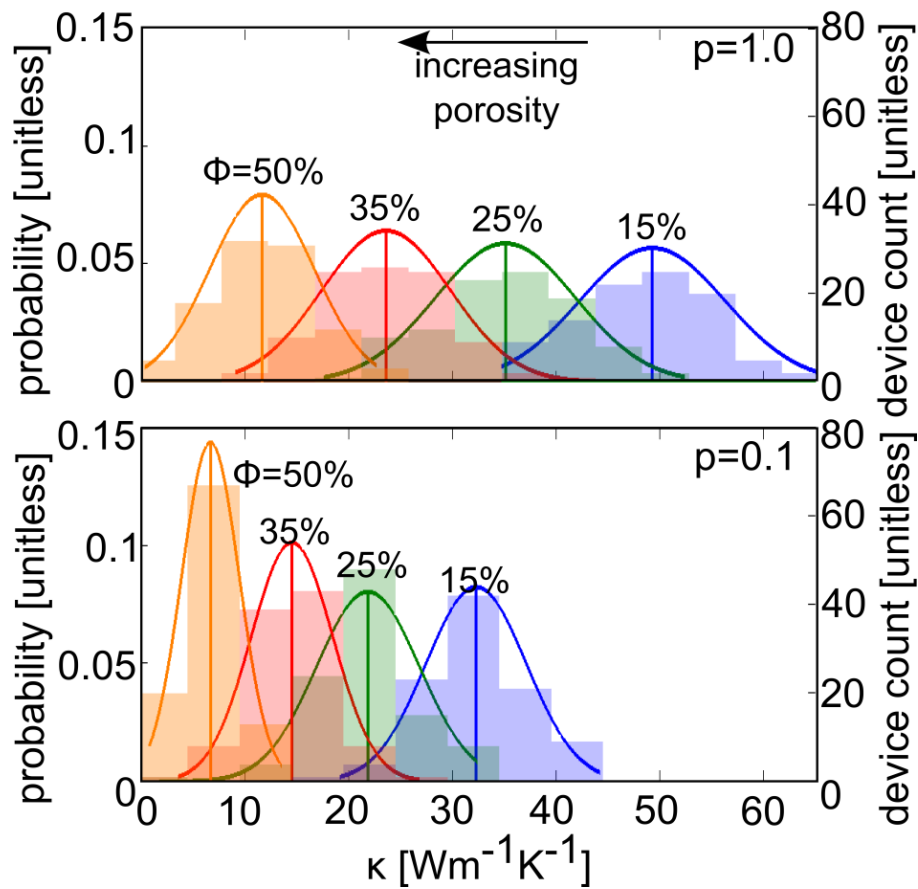

Figure 4 caption:

The thermal conductivity distributions in nanomeshes of different porosity. (a) Results for specular pore boundaries with p=1. (b) Results for near-diffusive pore boundaries with p=0.1. Results for porosity Φ=15% (blue), Φ=25% (green), Φ=35% (red), and Φ=50% (orange) are shown. 100 devices are simulated for each curve. The device count is indicated in the right y-axis. The arrow indicates the direction of increasing porosity.

Figure 5:

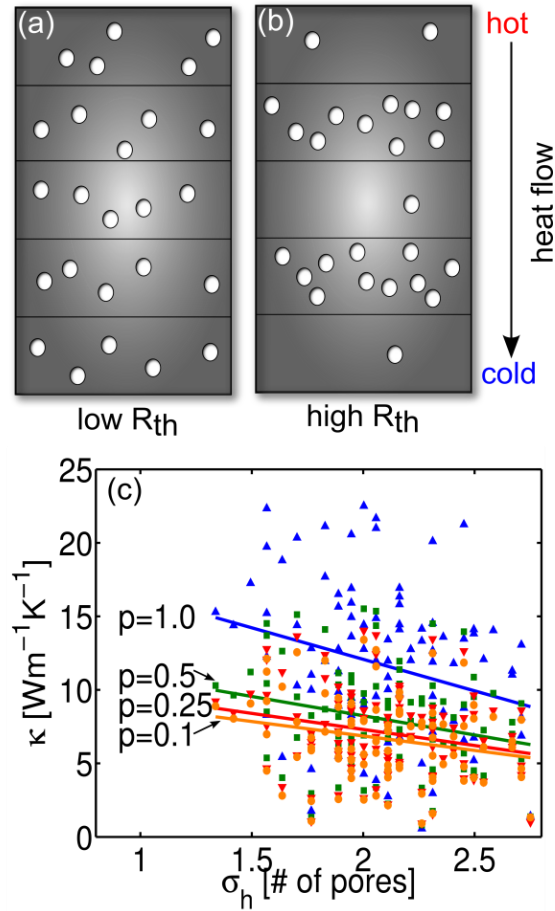

Figure 5 caption:

The influence of the non-uniform pore placement along the transport direction on the thermal conductivity. (a-b) Schematics showing cases in which the pores are placed uniformly in all segments of the channel along the transport direction (a), and non-uniformly (b). (c) The thermal conductivity versus the variation of pores $\sigma_h$ in the segments along the heat transport direction. Channels with porosity $\Phi=50\%$ are considered. Results for channels with boundary scattering specularity parameters p=1 (blue), p=0.5 (green), p=0.25 (red), p=0.1 (orange) are shown. 100 different geometries are simulated for each p. The symbols indicate the exact results. The lines present a linear least squares fit to the data points for each roughness case p.

23